\documentclass[prd,aps,nofootinbib,notitlepage,preprintnumbers]{revtex4}
\usepackage{graphicx,epsf,amsmath,amsfonts,amssymb,amsbsy}
\usepackage{epsfig}
\newcommand{\mat}{\left ( \begin{array}}
\newcommand{\emat}{\end{array} \right )}
\newcommand{\vect}{\left ( \begin{array}{c}}
\newcommand{\evect}{\end{array} \right )}

\def\cP{\mathcal P}

\begin{document}


\title{Hartree-Fock approach to dynamical mass generation in the generalized (2+1)-dimensional Thirring model}
\author{T. G. Khunjua $^{1}$, K. G. Klimenko $^{2}$, and R. N. Zhokhov $^{2,3}$ }

\affiliation{$^{1}$ The University of Georgia, GE-0171 Tbilisi, Georgia}
\affiliation{$^{2}$ State Research Center
of Russian Federation -- Institute for High Energy Physics,
NRC "Kurchatov Institute", 142281 Protvino, Moscow Region, Russia}
\affiliation{$^{3}$  Pushkov Institute of Terrestrial Magnetism, Ionosphere and Radiowave Propagation (IZMIRAN),
108840 Troitsk, Moscow, Russia}

\begin{abstract}
The (2+1)-dimensional generalized massless Thirring model with 4-component Fermi-fields is investigated by the 
Hartree-Fock method. The Lagrangian of this model is constructed from two different four-fermion structures.
One of them takes into account the vector$\times$vector channel of fermion interaction with coupling constant
$G_v$, the other - the scalar$\times$scalar channel with coupling $G_s$. At some relation between bare couplings 
$G_s$ and $G_v$, the Hartree-Fock equation for self-energy of fermions can be renormalized, and dynamical generation 
of the Dirac and Haldane fermion masses is possible. As a result, phase portrait of the model consists of two 
nontrivial phases. In the first one the chiral symmetry is spontaneously broken due to dynamical appearing of the 
Dirac mass term, while in the second phase a spontaneous breaking of 
the spatial parity $\cP$ is induced by Haldane mass term. It is shown that in the particular case of pure Thirring model, i.e. at $G_s=0$, the 
ground state of the system is indeed a mixture of these phases. Moreover, it was found that dynamical generation of 
fermion masses is possible for any finite number of Fermi-fields.
\end{abstract}
\maketitle
\section{Introduction}

Over the past few decades, in quantum field theory, much attention has been paid to the study of various
models in (2+1)-dimensional spacetime. This is due to the fact that in condensed matter physics there exist
quite a few phenomena (quantum Hall effect, high-temperature superconductivity, physical
processes in graphene, etc.) having a planar nature, 
and for their effective description it is convenient to use 
relativistic (2+1)-dimensional (D) models with four-fermion interaction. Among them is the Gross-Neveu model
\cite{Semenoff,Shovkovy,Gusynin,Vshivtsev,Khudyakov,Ebert,Gomes3}, Thirring model
\cite{Gomes,Hong,Rossini,Itoh,Hyun,Ahn,Gies,Janssen,Wipf,Wipf2,Hands2,Hands3,Hands4,Hands5} etc. Note that both models are 
renormalizable, at least in the framework of large-$N$ technique \cite{Krasnikov,Rosenstein,Hands} ($N$ is the number of 
fermion fields). At the same time, various nonperturbative 
approaches (such as $1/N$ expansion method, Gaussian effective potential and optimized expansion techniques 
\cite{Rosenstein,Klimenko,Kneur}, etc.) to the study of the massless (2+1)-D Gross-Neveu model predict its 
qualitatively identical properties (dynamic generation of fermion masses, spontaneous chiral symmetry breaking, etc.). 
However, using different methods for studying the massless (2+1)-D Thirring model built on the basis of a 
{\it four-component} reducible spinor representation for fermion fields,
leads to conflicting results. Indeed, in a number of papers (see, for example, \cite{Hong,Itoh}), this model was
investigated by the $1/N$ expansion method, where it was shown that only the Dirac fermion mass that breaks chiral symmetry 
can be generated (in this case, the spatial parity $\cP$ remains unbroken). In contrast,
an application of other research methods \cite{Rossini,Ahn} to the same Thirring model gives the opposite
result, because in these papers, the possibility of dynamical generation of a $\cP$-odd (but chirally invariant)
Haldane fermion mass was established. In addition, in the literature there is also a discrepancy in the predictions of the 
number of fermion fields $N$, with which dynamic mass generation is possible. Thus, for example, in the first of 
papers \cite{Hong,Itoh} this effect is predicted for any finite value of $N$, and in the second one only for 
$N<N_c=128/3\pi^2$, and so on. (More details about the inconsistency of the results of the study of the 
(2+1)-D Thirring model by various nonperturbative methods can be found, for example, in Refs. 
\cite{Janssen,Wipf,Wipf2,Hands2}.)

The discrepancy in the results points to the need for further more thorough study of the (2+1)-dimensional 
Thirring model both within the framework of well-known methods, as well as by attracting new approaches.

To this end, in the present paper we use the so-called Hartree-Fock (HF) approach to investigate the possibility of  
dynamical fermion mass generation in the (generalized) (2+1)-dimensional Thirring model with {\it four-component} spinors.
Earlier in Refs. \cite{Tamaz1,Tamaz2} it was used to study the (2+1)-D Gross-Neveu model. Wherein
it turned out that in the region of large $N$ the HF method predicts the same properties as the 
nonperturbative $1/N$-expansion method widely used to study this model. In the region of small $N$, where the 
$1/N$-expansion method is not applicable, the HF approach predicts the existence of other nontrivial phases of 
the three-dimensional Gross-Neveu model, including the spontaneously non-Hermitian phase of the model 
\cite{Tamaz2}. The essence of the HF method consists, firstly, in the using of the 
Cornwall-Jackiw-Tomboulis (CJT) effective action for composite operators $\Gamma(S)$ \cite{CJT} in field theory 
models with four-fermion interaction (here $S$ -- full fermion propagator satisfying the stationarity equation 
$\delta\Gamma/\delta S=0$), and, secondly, that $\Gamma(S)$ is considered in the first order in the coupling constants.
The resulting stationary equation takes the form of the well-known Hartree--Fock equation for fermion self-energy 
operator \cite{Buballa,Klevansky}. (This is the reason why we call this approach the HF method.) It should be especially 
noted that when studying the Thirring model by HF method, auxiliary vector fields are not used at all, as it is usually
practiced in most of the earlier approaches to the model, and due to which the mechanism of fermion mass generation  
in this model is more similar to the one found in (2+1)-D quantum electrodynamics (see, e.g., in Refs. 
\cite{Gomes,Itoh}).

In our work, based on the HF approach, we explore the properties of not only the pure massless (2+1)-D Thirring model 
(with single vector$\times$vector coupling $G_v$) composed of $N$ reducible four-component spinors, but also a more 
general model, invariant under the same continuous symmetry group, in which the Lagrangian contains an additional 
scalar$\times$scalar fermion interaction term with coupling constant $G_s$.
We show that, depending on the relationship between $G_v$ and $G_s$, the ground state of the generalized 
massless (2+1)-D Thirring model corresponds to either a chirally broken phase or phase in which fermions have a parity 
$\cP$ violating mass. In contrast, at $G_s=0$ in the ground state of the pure (2+1)-dimensional massless Thirring 
model these phases can coexist. Moreover, it is clear from our HF consideration that dynamical generation of fermion 
mass is allowed to occur at any finite value of $N$. 

The paper is organized as follows. In section \ref{IIA} we present the $N$-flavor massless 
(2+1)-D generalized Thirring model symmetric under discrete chiral and spatial $\cP$ reflections. Here it is also 
shown that model is invariant under continuous $U(2N)$ group, and two different fermion mass terms, Dirac and 
Haldane, are defined.  
In section \ref{IIB} the CJT effective action $\Gamma(S)$ of the composite bilocal and bifermion operator 
$\overline\psi (x)\psi (y)$ is constructed, which is actually the functional of the full fermion propagator $S(x,y)$. 
Then, the unrenormalized expression for $\Gamma(S)$ is obtained up to a first order 
in the bare coupling constants $G_{s,v}$ (it is the so-called Hartree-Fock approximation). Based on this expression, 
we show in section III  that for a some
well-defined behavior of the bare coupling constants $G_{s,v}(\Lambda)$ vs cutoff parameter $\Lambda$, there exist two different
renormalized, i.e. without ultraviolet divergences, solutions of the stationary Hartree-Fock equation for 
the propagator. One of them corresponds to a phase in which the Haldane fermion mass term arises dynamically, and 
parity $\cP$ is spontaneously broken down. Another solution of the HF equation corresponds to a 
chiral symmetry breaking phase with dynamically emerging Dirac mass term.  
Finally, in section IV we use the renormalization group formalism and show that in the plane of dimensionless 
coupling constants there is at least one ultraviolet-stable fixed point of the model. Appendix A contains 
some information about two- and four-dimensional spinor representations of the $SO(2,1)$ group, whereas Appendix B 
gives all details of calculating the effective action $\Gamma(S)$ in the HF approximation. 

\section{(2+1)-dimensional generalized Thirring model and Hartree-Fock approach}

\subsection{The model, its symmetries, etc}\label{IIA}

The Lagrangian of the generalized massless and $N$-flavored (2+1)-D Thirring model under 
consideration has the following form (see, e.g., in Refs. \cite{Mesterhazy,Gies})
\begin{eqnarray}
 L=\overline \Psi_k\gamma^\nu i\partial_\nu \Psi_k&-& \frac {G_v}{2N}\left
(\overline \Psi_k\gamma^\mu\Psi_k\right )\left (\overline \Psi_k\gamma_\mu\Psi_k\right )+
\frac {G_s}{2N}\left (\overline \Psi_k\tau\Psi_k\right )^2,
\label{t1}
\end{eqnarray}
where for each $k=1,...,N$ the field $\Psi_k\equiv \Psi_k(t,x,y)$ is a (reducible) four-component Dirac spinor
(its spinor indices are omitted in Eq. (\ref{t1})), $\gamma^\nu$ 
($\nu=0,1,2$) are 4$\times$4 matrices acting in the four-dimensional spinor space (the algebra
of these $\gamma$-matrices and their particular representation used in the present paper
is given in Appendix \ref{ApC}, where in addition the matrices $\gamma^3,\gamma^5$ and $\tau=-i\gamma^3\gamma^5$
are also introduced), and the summation over repeated $k$- and $\mu,\nu$-indices is assumed in Eq. 
(\ref{t1}) and below. The bare coupling constants $G_v$ and $G_s$ have a dimension of [mass]$^{-1}$. As discussed,
e.g., in Refs. \cite{Gies,Mesterhazy}, at $N=2$ the model (\ref{t1}) provides a fairly 
good description of the low-energy physics of graphene in the continuum limit. But  
we consider the $N$-flavor variant of the model in order to compare its phase structure 
obtained in the framework of the HF effective approach with the results of the large-$N$ 
investigation \cite{Hong,Itoh}. 

Together, all four-component spinor fields $\Psi_k$ ($k=1,...,N$) form a fundamental multiplet of the $U(N)$ group, so the invariance of the 
Lagrangian (\ref{t1}) with respect to this group is obvious (and in the following the $U(N)$-symmetry of the model 
remains unbroken). It is not so obvious that, in reality, the continuous symmetry group of the three-dimensional 
generalized Thirring 
model is wider and is $U(2N)$. This fact can be easily established if we rewrite the expression (\ref{t1}) in terms 
of two-component spinors. Namely, for each fixed $k=1,...,N$ we set $\Psi_k^t=(\psi_{2k-1}^t,
\psi_{2k}^t)$, where the symbol $^t$ means the transposition operation, and $\psi_{2k-1}$ and $\psi_{2k}$ are 
two-component spinors (see in Appendix \ref{ApC}). Then we have:
\begin{eqnarray}
 &&L_0\equiv \overline \Psi_k\gamma^\nu i\partial_\nu \Psi_k=\overline \psi_1\tilde\gamma^\nu i\partial_\nu \psi_1+
 \overline \psi_2\tilde\gamma^\nu i\partial_\nu \psi_2+\cdots +\overline \psi_{2N}\tilde\gamma^\nu i\partial_\nu
 \psi_{2N},\nonumber\\
 &&~~~~~~~~~~\overline \Psi_k\gamma^\nu\Psi_k=\overline \psi_1\tilde\gamma^\nu \psi_1+
 \overline \psi_2\tilde\gamma^\nu \psi_2+\cdots +\overline \psi_{2N}\tilde\gamma^\nu \psi_{2N},\nonumber\\
&&~~~~~~~~~~~~~~~\overline \Psi_k\tau\Psi_k=\overline \psi_1 \psi_1+
 \overline \psi_2 \psi_2+\cdots +\overline \psi_{2N} \psi_{2N}, 
\label{t2}
\end{eqnarray}
where $\tilde\gamma^\nu$ are $2\times 2$ matrices (see in Appendix A). Assuming formally that the set of all
two-component spinors $\psi_{2k-1}$ and $\psi_{2k}$ ($k=1,..,N)$ forms a fundamental representation of the 
$U(2N)$ group, it is easy to see that both the structures (\ref{t2}) and the entire Lagrangian (\ref{t1}) are 
invariant under this group.

More important for us is that the 
Lagrangian (\ref{t1}) is invariant under three discrete transformations, two of them are the so-called 
chiral transformations $\Gamma^5$ and $\Gamma^3$,
\begin{eqnarray}
 \Gamma^5:&~~&\Psi_k(t,x,y)\to  \gamma^5\Psi_k(t,x,y);~~
 \overline\Psi_k(t,x,y)\to  -\overline\Psi_k(t,x,y)\gamma^5,\nonumber\\
 \Gamma^3:&~~&\Psi_k(t,x,y)\to  \gamma^3\Psi_k(t,x,y);~~
 \overline\Psi_k(t,x,y)\to  -\overline\Psi_k(t,x,y)\gamma^3.
 \label{n4}
\end{eqnarray}
The rest one is the space reflection, or parity, transformation $\cP$ under which $(t,x,y)\to 
(t,-x,y)$ \footnote{In 2+1 dimensions, parity corresponds to inverting only one 
spatial axis \cite{Semenoff,Appelquist}, since the inversion of both axes is equivalent to 
rotating the entire space by $\pi$.} and
\begin{eqnarray}
 \cP:~~\Psi_k(t,x,y)\to  \gamma^5\gamma^1\Psi_k(t,-x,y);~~\overline\Psi_k(t,x,y)\to 
 \overline\Psi_k(t,-x,y)\gamma^5\gamma^1.
 \label{nn4}
\end{eqnarray}
Due to the  symmetry of the model (\ref{t1}) with respect to each of the discrete 
$\Gamma^5$, $\Gamma^3$ and $\cP$ transformations, different mass 
terms are prohibited to appear perturbatively in this Lagrangian. Indeed, the most popular Dirac mass term has the 
form $m_D\overline \Psi_k \Psi_k=m_D(\overline\psi_{2k-1}\psi_{2k-1}
-\overline\psi_{2k}\psi_{2k})$, but it breaks 
both $U(2N)$ and chiral $\Gamma^5$ and $\Gamma^3$ symmetries of the model, although it is $\cP$-even. 
There is another well-known fermionic mass term that is often discussed in the literature. This is a mass term of 
the form $m_H\overline \Psi_k\tau \Psi_k=m_H(\overline\psi_{2k-1}\psi_{2k-1}+\overline\psi_{2k}\psi_{2k})$ 
(recall, here the 4$\times$4 matrix $\tau$ is defined 
in Appendix \ref{ApC}) and sometimes it is refered to as the Haldane mass term (see, e.g., in Refs. 
\cite{Ebert,Hands3}). \footnote{The appearance of the Haldane mass term is related 
to the parity anomaly in (2+1) dimensions, to generation of the Chern-Simons topological mass 
of gauge fields \cite{Klimenko3,Gomes2}, as well as to the integer quantum Hall effect in planar 
condensed matter systems without external magnetic field, etc \cite{Haldane}.} 
But nonzero Haldane mass $m_H$ breaks the parity $\cP$ invariance of the model (although it is $U(2N)$ invariant and
chirally $\Gamma^5$ and $\Gamma^3$ symmetric). So both Dirac and Haldane mass terms cannot appear in the model
(\ref{t1}) when it is studied by the usual perturbative technique. However, within a framework of nonperturbative 
approximations (for example, in the $1/N$ expansion, etc), fermion mass can arise dynamically, thereby breaking the 
original symmetry in a spontaneous way. 

In our paper, we continue the investigation of (2+1)-D models with four-fermion interactions by the 
so-called HF method, which was started in our papers  \cite{Tamaz1,Tamaz2}. This time we use it to explore the possibility
of dynamical mass generation within the framework of the generalized Thirring model (\ref{t1}). Note that 
theoretical ground of the HF method is the effective Cornwall-Jakiw-Tomboulis (CJT) action for 
composite operators \cite{CJT}, which also provides a systematic way to go beyond the HF approximation.

\subsection{From CJT to Hartree-Fock approach}\label{IIB}

Let us define $Z(K)$, the generating functional of the Green's functions of bilocal 
fermion-antifermion composite operators $\sum_{k=1}^N\overline\Psi_k^\alpha(x)\Psi_{k\beta}(y)$ 
in the framework of the (2+1)-D Thirring model (\ref{t1}) (the corresponding technique 
for theories with four-fermion interaction is elaborated in details, e.g., in Ref. \cite{Rochev}) 
\begin{eqnarray}
 Z(K)\equiv\exp(iNW(K))=\int {\cal D}\overline\Psi_k {\cal D}\Psi_k \exp\Big
(i\Big [ I(\overline\Psi,\Psi)+\int d^3xd^3y\overline\Psi_k^\alpha(x)K_\alpha^\beta(x,y)
\Psi_{k\beta}(y) \Big ]\Big ),
 \label{36}
\end{eqnarray}
where $\alpha,\beta =1,2,3,4$ are spinor indices, $K_\alpha^\beta(x,y)$ is a bilocal source of 
the fermion bilinear composite field $\overline\Psi_k^\alpha(x)\Psi_{k\beta}(y)$ (recall that in 
all expressions the summation over repeated indices is assumed). 
\footnote{We denote a matrix element of an arbitrary matrix (operator) $\hat A$ acting in the 
four dimensional spinor space by the symbol $A^\alpha_\beta$, where the upper (low) index 
$\alpha $($\beta$) is the column (row) number of the matrix $\hat A$. In particular, the matrix 
elements of any $\gamma^\mu$ matrix is denoted by $(\gamma^\mu)^\alpha_\beta$.}
Moreover, $I(\overline\Psi,\Psi)=\int Ld^3x$, where $L$ is the Lagrangian (\ref{t1}) of the model under consideration. Hence,
$$
I(\overline\Psi ,\Psi)=\int d^3xd^3y\overline\Psi_k^\alpha(x)D_\alpha^\beta(x,y)\Psi_{k\beta} (y)+
I_{int}(\overline\Psi_k^\alpha\Psi_{k\beta}),~~D_\alpha^\beta(x,y)=
\left(\gamma^\nu\right)_\alpha^\beta i\partial_\nu\delta^3(x-y),~~I_{int}=I_{v}+I_{s},
$$\vspace{-0.5cm}
\begin{eqnarray}
I_{v}&=& -\frac {G_v}{2N}\int d^3x\left (\overline \Psi_k\gamma^\mu\Psi_k\right )
\left (\overline \Psi_l\gamma_\mu\Psi_l\right )\nonumber\\
&=&-\frac {G_v}{2N}\int d^3xd^3td^3ud^3v\delta^3 (x-t)\delta^3 (t-u)\delta^3 (u-v) \overline
\Psi_k^\alpha(x)(\gamma^\mu)_\alpha^\beta\Psi_{k\beta}(t) \overline\Psi_l^\rho(u)
(\gamma_\mu)_\rho^\xi\Psi_{l\xi}(v),\nonumber\\
I_{s}&=& \frac {G_s}{2N}\int d^3x\left (\overline \Psi_k\tau\Psi_k\right )
\left (\overline \Psi_l\tau\Psi_l\right )\nonumber\\
&=&\frac {G_s}{2N}\int d^3xd^3td^3ud^3v\delta^3 (x-t)\delta^3 (t-u)\delta^3 (u-v) \overline
\Psi_k^\alpha(x)(\tau)_\alpha^\beta\Psi_{k\beta}(t) \overline\Psi_l^\rho(u)
(\tau)_\rho^\xi\Psi_{l\xi}(v).
 \label{360}
\end{eqnarray}
Note that in Eq. (\ref{360}) and similar expressions below, $\delta^3(x-y)$ denotes the 
three-dimensional Dirac delta function. There is an alternative expression for $Z(K)$,
\begin{eqnarray}
Z(K)&=&\exp\Big (iI_{int}\Big (-i\frac{\delta}{\delta K}\Big )\Big )\int 
{\cal D}\overline\Psi_k {\cal D}\Psi_k \exp\Big (i
\int d^3xd^3y\overline\Psi_k(x)\Big [D(x,y)+K(x,y)\Big ]\Psi_k (y)\Big )
\nonumber\\&=&\exp\Big (iI_{int}\Big (-i\frac{\delta}{\delta K}\Big )\Big )\Big 
[\det\big (D(x,y)+K(x,y)\big )\Big ]^N\nonumber\\&=&\exp\Big (iI_{int}\Big 
(-i\frac{\delta}{\delta K}\Big )\Big )\exp \Big [N{\rm Tr}\ln \big (D(x,y)+K(x,y)\big )\Big ],
\label{036}
\end{eqnarray}
where instead of each bilinear form $\overline\Psi_k^\alpha(s)\Psi_{k\beta}(t)$ appearing in $I_{int}$ of
Eq. (\ref{360}) we use a variational derivative $-i\delta /\delta K^\beta_\alpha (s,t)$.
Moreover, the Tr-operation in Eq. (\ref{036}) means the trace both over spacetime and spinor coordinates. 
The effective action (or CJT effective action) of the composite bilocal and bispinor operator 
$\overline\Psi_k^\alpha(x)\Psi_{k\beta}(y)$ is defined as a functional $\Gamma (S)$ of the full 
fermion propagator $S^\alpha_\beta(x,y)$ by a Legendre transformation of the functional 
$W(K)$ entering in Eq. (\ref{36}),
\begin{eqnarray}
\Gamma (S)=W(K)-\int d^3xd^3y S^\alpha_\beta(x,y)K_\alpha^\beta(y,x),
\label{0360}
\end{eqnarray}
where
\begin{eqnarray}
S^\alpha_\beta(x,y)=\frac{\delta W(K)}{\delta K_\alpha^\beta(y,x)}.
 \label{37}
\end{eqnarray}
Taking into account the relation (\ref{36}), it is clear that $S(x,y)$ is the full fermion 
propagator at $K(x,y)=0$. Hence, in order to construct the CJT effective action $\Gamma (S)$ of Eq. 
(\ref{0360}), it is necessary to solve Eq. (\ref{37}) with respect to $K$ and then to use the obtained expression 
for $K$ (in fact, it is a functional of $S$) in Eq. (\ref{0360}). It follows from the definition (\ref{0360})-(\ref{37}) that
\begin{eqnarray}
\frac{\delta\Gamma (S)}{\delta S^\alpha_\beta(x,y)}=\int d^3ud^3v\frac{\delta W(K)}{\delta K^\mu_\nu(u,v)}\frac{\delta K^\mu_\nu(u,v)}{\delta S^\alpha_\beta(x,y)}-K_\alpha^\beta(y,x)-\int d^3ud^3v S_\mu^\nu(v,u)\frac{\delta K^\mu_\nu(u,v)}{\delta S^\alpha_\beta(x,y)}.
 \label{037}
\end{eqnarray}
(In Eq. (\ref{037}) and below, the Greek letters $\alpha,\beta,\mu,\nu,$ etc, also denote
the spinor indices, i.e. $\alpha,...\nu,...=1,...,4$.) Now, due to the relation (\ref{37}), it is easy to see that 
the first term in Eq. (\ref{037}) cansels there the last term, so
\begin{eqnarray}
\frac{\delta\Gamma (S)}{\delta S^\alpha_\beta(x,y)}=-K_\alpha^\beta(y,x).
\label{370}
\end{eqnarray}
Hence, in the true theory, in which bilocal sources $K_\alpha^\beta(y,x)$ are zero, the full fermion propagator 
is a solution of the following stationary equation,
\begin{eqnarray}
\frac{\delta\Gamma (S)}{\delta S^\alpha_\beta(x,y)}=0.
\label{0370}
\end{eqnarray}
Note that in the nonperturbative CJT approach the stationary/gap equation (\ref{0370}) 
for fermion propagator $S^\beta_\alpha(x,y)$  is indeed a Schwinger--Dyson equation \cite{Rochev}.
Further, in order to simplify the calculations and obtain a more detailed information about 
the phase structure of the model, we calculate both the effective action (\ref{0360}) and the gap equation 
(\ref{0370}) up to a first order in the couplings $G_v$ and $G_s$. 

We call such an approach to studying the properties of any 
model with four-fermion interactions (including the generalized Thirring model (\ref{t1})) as the 
Hartree-Fock method (a more detailed justification for this name is given at the end of this section).

In this case (see in Appendix \ref{ApB})
\begin{eqnarray}
&&\Gamma (S)=i{\rm Tr}\ln \big (iS\big )+\int d^3xd^3y S^\alpha_\beta(x,y)D_\alpha^\beta(y,x)
-\frac{G_v}2\int d^3s~ {\rm tr}\big [\gamma^\rho S(s,s)
\big ] {\rm tr}\big [\gamma_\rho S(s,s)\big ]
\nonumber\\
&&+\frac{G_v}{2N}\int d^3s~ {\rm tr}\Big [\gamma^\rho S(s,s)\gamma_\rho S(s,s)\Big ]
+\frac{G_s}2\int d^3s \left ({\rm tr}\big [\tau S(s,s)\big ]\right )^2
-\frac{G_s}{2N}\int d^3s~ {\rm tr}\Big [\tau S(s,s)\tau S(s,s)\Big ].
 \label{n420}
\end{eqnarray}
Notice that in Eq. (\ref{n420}) the symbol ${\rm tr}$ means the trace of an operator 
over spinor indices only, but the symbol ${\rm Tr}$ is still the trace operation both over spacetime coordinates and spinor 
indices. Moreover, the expression for operator $D(x,y)$ is presented in Eq. (\ref{360}). The stationary 
equation (\ref{0370}) for the CJT effective action (\ref{n420}) looks like \footnote{The first 
term on the left-hand side of Eq. (\ref{n0420}) can be easily obtained using Eq. (\ref{A50})
from Appendix \ref{ApB}.} 
\begin{eqnarray}
-i\Big [S^{-1}\Big ]^\beta_\alpha(x,y)-D_\alpha^\beta(x,y)&=&G_s\tau^\beta_\alpha{\rm tr}
\big [\tau S(x,y)\big ]\delta^3(x-y)-G_v\big (\gamma^\rho\big )^\beta_\alpha{\rm tr}
\big [\gamma_\rho S(x,y)\big ]\delta^3(x-y)\nonumber\\
&-&\frac{G_s}N \big [\tau S(x,y)\tau\big ]^\beta_\alpha
\delta^3 (x-y)+\frac{G_v}N \big [\gamma^\rho S(x,y)\gamma_\rho\big ]^\beta_\alpha\delta^3 (x-y).
\label{n0420}
\end{eqnarray}
Now suppose that $S(x,y)$ is a translationary invariant operator. Then 
\begin{eqnarray}
S^\beta_\alpha(x,y)\equiv S^\beta_\alpha(z)&=&\int\frac{d^3p}{(2\pi)^3}
\overline{S^\beta_\alpha}(p)e^{-ipz},~~~\overline{S^\beta_\alpha}(p)=\int d^3z S^\beta_\alpha(z)e^{ipz},\nonumber\\
\Big (S^{-1}\Big )^\beta_\alpha(x,y)&\equiv& \Big (S^{-1}\Big )^\beta_\alpha(z)=\int\frac{d^3p}{(2\pi)^3}
\overline{(S^{-1})^\beta_\alpha}(p)e^{-ipz},
\label{n43}
\end{eqnarray}
where $z=x-y$ and $\overline{S^\beta_\alpha}(p)$ is a Fourier transformation of $ S^\beta_\alpha(z)$.
After Fourier transformation the Eq. (\ref{n0420}) takes the form
\begin{eqnarray}
-i\overline{(S^{-1})^\beta_\alpha}(p)- (\hat p)^\beta_\alpha&=&
G_s\tau^\beta_\alpha\int\frac{d^3q}{(2\pi)^3}~{\rm tr}\big [\tau\overline{S}(q)\big ]
-G_v\big (\gamma^\rho\big )^\beta_\alpha\int\frac{d^3q}{(2\pi)^3}~{\rm tr}
\big [\gamma_\rho\overline{S}(q)\big ]\nonumber\\
&-&\frac{G_s}N \int\frac{d^3q}{(2\pi)^3}~\big [\tau\overline{S}(q)\tau\big ]^\beta_\alpha+
\frac{G_v}N \int\frac{d^3q}{(2\pi)^3}~\big [\gamma^\rho\overline{S}(q)\gamma_\rho\big ]^\beta_\alpha,
 \label{n043}
\end{eqnarray}
where $\hat p=p_\nu\gamma^\nu$. It is clear from Eq. (\ref{n043}) that in the framework of the four-fermion model 
(\ref{t1}) the Schwinger-Dyson equation for fermion propagator $\overline{S}(p)$ reads in the first order
in $G_{s,v}$ like the Hartree-Fock equation for its self-energy operator $\Sigma(p)$ (the last quantity is nothing 
but the expression on the left side of this equation). As a result, we will henceforth refer to Eq. 
(\ref{n043}) as the Hartree-Fock equation. In particular, 
the first two terms on the right-hand side of Eq. (\ref{n043}) are the 
so-called Hartree contribution, whereas the last two terms there are the Fock contribution
to fermion self energy (for details, see, e.g., the section 4.3.1 in Ref. \cite{Buballa} or the section II C in 
Ref. \cite{Klevansky}). 

Finally, note that both the CJT (or HF) effective action (\ref{n420}) and its stationary equation 
(\ref{n0420})-(\ref{n043}), in which $G_{s,v}$ are bare coupling constants, contain ultraviolet (UV)
divergences and need to be renormalized. In the next sections, using a rather general ansatz for 
propagator $\overline{S}(p)$, we find the corresponding modes of the coupling constants $G_{s,v}$ 
behavior vs cutoff parameter $\Lambda$, such that  there occurs a renormalization of the gap Hartree-Fock equation 
(\ref{n043}), and it is possible to obtain its finite solution in the limit $\Lambda\to\infty$.

\section{Possibility for dynamical generation of the Dirac and Haldane masses }

Let us study on the basis of the HF equation (\ref{n043}) the possibility for 
dynamical generation of the Hermitian mass term $\overline \Psi_k(m_D+m_H\tau)\Psi_k$ in the massless 
(2+1)-D Thirring model (1). It means that we should find the solution $\overline S(p)$ of this equation, 
which looks like
\begin{eqnarray}
&&\overline {S}(p)=-i\left (\hat p+m_D+m_H\tau\right )^{-1}= -i  \left (\begin{array}{cc}
\tilde p+m_D+m_H, & 0\\
0~~, &-\tilde p+m_D-m_H
\end{array}\right )^{-1}\nonumber\\
&=&
-i  \left (\begin{array}{cc}
(\tilde p+m_D+m_H)^{-1}, & 0\\
0~~, &(-\tilde p+m_D-m_H)^{-1}
\end{array}\right )=
-i  \left (\begin{array}{cc}
\frac{\tilde p-m_D-m_H}{p^2-(m_D+m_H)^2}, & 0\\
0~~, &\frac{-\tilde p-m_D+m_H}{p^2-(m_D-m_H)^2}
\end{array}\right )
, \label{m3}
\end{eqnarray}
where $m_D$ and $m_H$ are finite unknown quantities, and in Eq. (\ref{m3}) the 4$\times$4 matrix 
$\overline {S}(p)$ is presented in the form of a 2$\times$2 matrix each element of which is, in tern, 
a 2$\times$2 matrix. Moreover, there $\tilde p=\tilde \gamma^\nu p_\nu$, where $\tilde\gamma^\nu$ 
($\nu=0,1,2$) are 2$\times$2 Dirac gamma-matrices (see in Appendix \ref{ApC}). It is evident that in this case
$\overline{S^{-1}}(p)=i(\hat p+m_D+\tau m_H)$. Using Eq. (\ref{m3}) in the HF gap equation (\ref{n043}),
we obtain for the quantities $m_D$ and $m_H$ the following {\it unrenormalized} system of gap equations
\begin{eqnarray}
m_D&=&\left (\frac{3iG_v}{2N}-\frac{iG_s}{2N}\right )\int\frac{d^3p}{(2\pi)^3}\left\{
\frac{m_D+m_H}{p^2-(m_D+m_H)^2}+\frac{m_D-m_H}{p^2-(m_D-m_H)^2}\right\},\nonumber\\
m_H&=&\left (2iG_s-\frac{iG_s}{2N}+\frac{3iG_v}{2N}\right )\int\frac{d^3p}{(2\pi)^3}\left\{
\frac{m_D+m_H}{p^2-(m_D+m_H)^2}-\frac{m_D-m_H}{p^2-(m_D-m_H)^2}\right\}.\label{m4}
\end{eqnarray}
Performing in the integrals of Eq. (\ref{m4}) a Wick rotation, $p_0\to i p_3$, and then using in the obtained 
three-dimensional Euclidean integration space the spherical coordinate system, $p_3=p\cos\theta, 
p_1=p\sin\theta\cos\phi, p_2=p\sin\theta\sin\phi$, we have (after integration over angles, $0\le\theta\le\pi, 
0\le\phi\le2\pi$, and cutting off the region of integration of the variable $p$, $0\le p\le \Lambda$) 
the following {\it regularized} gap system
\begin{eqnarray}
m_D&=&\left (\frac{3G_v}{2N}-\frac{G_s}{2N}\right )\int_0^\Lambda\frac{p^2~dp}{2\pi^2}\left\{
\frac{m_D+m_H}{p^2+(m_D+m_H)^2}+\frac{m_D-m_H}{p^2+(m_D-m_H)^2}\right\},\nonumber\\
m_H&=&\left (2G_s-\frac{G_s}{2N}+\frac{3G_v}{2N}\right )\int_0^\Lambda\frac{p^2~dp}{2\pi^2}\left\{
\frac{m_D+m_H}{p^2+(m_D+m_H)^2}-\frac{m_D-m_H}{p^2+(m_D-m_H)^2}\right\}.\label{m7}
\end{eqnarray}
Since
\begin{eqnarray}
\int_0^\Lambda\frac{p^2}{p^2+M^2}dp=\Lambda-\frac \pi 2 |M|+
M{\cal O}\left (\frac M\Lambda\right ),
 \label{m6}
\end{eqnarray}
the equations (\ref{m7}) can be presented in the following asymptotic forms
\begin{eqnarray}
\frac{m_D}{A}&=&2m_D\Lambda-\frac{\pi}{2}\big[ (m_D+m_H)|m_D+m_H|+(m_D-m_H)|m_D-m_H|\big ]
+m_D{\cal O}\left (\frac{m_D}\Lambda\right ),\nonumber\\
\frac{m_H}{B}&=&2m_H\Lambda-\frac{\pi}{2}\big[ (m_D+m_H)|m_D+m_H|-(m_D-m_H)|m_D-m_H|\big ]
+m_H{\cal O}\left (\frac{m_H}\Lambda\right ),\label{m71}
\end{eqnarray}
where 
\begin{eqnarray}
&&A=\frac{3G_v}{4N\pi^2}-\frac{G_s}{4N\pi^2},~~
B=\frac{G_s}{\pi^2}-\frac{G_s}{4N\pi^2}+\frac{3G_v}{4N\pi^2}.
\label{m72}
\end{eqnarray}
To remove the UV divergences from Eqs. (\ref{m71}), we suppose that bare quantities $A\equiv A(\Lambda)$ and $B\equiv B(\Lambda)$ are such that
\begin{eqnarray}
\frac{1}{A(\Lambda)}&=&2\Lambda+\frac{\pi}{2}g_A+g_A{\cal O}\left (\frac{g_A}\Lambda\right ),\nonumber\\
\frac{1}{B(\Lambda)}&=&2\Lambda+\frac{\pi}{2}g_B+g_B{\cal O}\left (\frac{g_B}\Lambda\right ),
\label{m73}
\end{eqnarray}
where $g_A$ and $g_B$ are some finite $\Lambda$-independent and renormalization group invariant 
quantities with dimension of mass. In this case, at $\Lambda\to\infty$ the system of stationary
equations (\ref{m71}) aquire the following {\it renormalized } form
\begin{eqnarray}
&&m_Dg_A+ (m_D+m_H)|m_D+m_H|+(m_D-m_H)|m_D-m_H|=0,\nonumber\\
&&m_Hg_B+ (m_D+m_H)|m_D+m_H|-(m_D-m_H)|m_D-m_H|=0.\label{m74}
\end{eqnarray}
Moreover, it is clear from Eq. (\ref{m73}) that at sufficiently large values of $\Lambda$
\begin{eqnarray}
A(\Lambda)&=&\frac{1}{2\Lambda}\left(1-\frac{\pi}{4\Lambda}g_A+\cdots\right),\nonumber\\
B(\Lambda)&=&\frac{1}{2\Lambda}\left(1-\frac{\pi}{4\Lambda}g_B+\cdots\right).
\label{m740}
\end{eqnarray}
So, taking into account the relations (\ref{m72}), we have for the bare constants $G_{s,v}$ 
the following asymptotic expansions at $\Lambda\to\infty$  
\begin{eqnarray}
G_s\equiv G_s(\Lambda)&=&\pi^2 (B-A)
=\frac{\pi^3}{8\Lambda^2}(g_A-g_B)+\cdots,\nonumber\\
G_v\equiv G_v(\Lambda)&=&\frac{\pi^2}{3}(B-A)+\frac{4\pi^2N}{3}A=\frac{2\pi^2N}{3\Lambda}-
\frac{\pi^3}{24\Lambda^2}\big [(4N-1)g_A+g_B\big ]+\cdots.
\label{m75}
\end{eqnarray}

As a rule, the stationary equation (\ref{0370}) have several solutions.
To find which one is more preferable, it is necessary to consider the so-called 
CJT effective potential $V(S)$ of the model which is determined on the basis of the CJT effective action (\ref{0360}) 
by the following relation \cite{CJT}
\begin{eqnarray}
V(S)\int d^3x\equiv -\Gamma(S)\Big |_{\rm transl.-inv.~S(x,y) },
 \label{m76}
\end{eqnarray}
and that solution $S$ of the stationarity equation (\ref{0370}), on which the effective potential $V(S)$ takes the 
least value, will correspond to the true fermion propagator $S(x,y)$ of the model.
To find CJT effective potential $V(S)$ in the Hartree-Fock approximation, we should use in Eq. (\ref{m76}) the 
expressions (\ref{n420}) and (\ref{m3}) for CJT effective action $\Gamma(S)$ and for the full fermion propagator 
$S(x,y)$, respectively. But in this case the obtained expression for $V(S)$ contains UV-divergences. However,
they are eliminated if bare couplings $G_{s,v}$ are constrained by relations (\ref{m75}).  As a result, 
in the Hartree-Fock approximation we obtain for the CJT effective potential 
$V(S)\equiv V(m_H,m_D)$ the following {\it renormalized} expression  
\begin{eqnarray}
V(m_D,m_H)=\frac{1}{12\pi}\Big (3 g_Am_D^2+3 g_B m_H^2 +2|m_D+m_H|^3+
2 |m_D-m_H|^3\Big )
\label{m77}
\end{eqnarray}
(this expression is valid up to unessential $m_D,m_H$-independent infinite constant).
Note that the HF gap equations (\ref{m74}) are also the stationary equations for the effective 
potential (\ref{m77}). Now it is clear that the form of the global minimum point (GMP) of the function 
$V(m_D,m_H)$ determines the phase structure of the generalized Thirring model (\ref{t1}) when 
coupling constants $G_{s,v}$ are constrained by the conditions (\ref{m75}).

Let us study the behavior of the GMP of the function $V(m_D,m_H)$ (\ref{m77}) vs finite couplings $g_{A,B}$. First, 
note that this function is symmetric under the transformations $m_D\to -m_D$ and/or $m_H\to -m_H$. 
So, for simplicity, it is enough to look for its GMP only in the region $m_D,m_H\ge 0$. Second, it is evident 
that at $g_A, g_B\ge 0$ the GMP of $V(m_D,m_H)$ lies at the point $(m_D=0,m_H=0)$, which means that no fermion 
masses are generated in this region, and symmetry remains intact. In other
regions for $g_A$ and $g_B$, it is easy to find the following form of the GMP of the function $V(m_D,m_H)$ 
(\ref{m77}):

\underline{The region $g_B<0$, $g_A>g_B$.} In this case the system of gap equations (\ref{m74}) has a nontrivial 
solution of the form $(m_D=0,m_H=-g_B/2)$, which corresponds to the free energy density equals to 
$V(m_D=0,m_H=-g_B/2)=\frac{1}{48\pi}g_B^3<0$, and this quantity is smaller, than the value of the CJT effective 
potential (\ref{m77}) at another, 
trivial, solution $(m_D=0,m_H=0)$ of the gap equations (\ref{m74}). So, in the region under consideration only 
the Haldane mass term can be generated dynamically, and parity $\cP$ (\ref{nn4}) is broken spontaneously.

\underline{The region $g_A<0$, $g_B>g_A$.} In this case the GMP of $V(m_D,m_H)$ is arranged at the 
point $(m_D=-g_A/2,m_H=0)$. Hence, in this $g_A,g_B$-region only the Dirac mass is allowed to be generated
and the phase with spontaneous breaking both of the chiral (see in Eq. (\ref{n4})) and $U(2N)$ symmetries is 
realized (parity $\cP$ is conserved). The density of free energy in this phase is equal to 
$V(m_D=-g_A/2,m_H=0)=\frac{1}{48\pi}g_A^3<0$.

In terms of the finite $g_A,g_B$-couplings, the phase portrait of the model is depicted in Fig. 1. Note that on the line 
$L=\{(g_A,g_B):g_A=g_B, g_A<0\}$ of this figure there is a first-order phase transition from chiral symmetry broken (CSB)
(at $g_B>g_A$) to $\cP$-broken phase (at $g_A>g_B$). On the line $L$ we have an equality of the free energy densities
of the ground states of these phases. In other words, it means that at $g_A=g_B$ the two phases coexist. 
Moreover, it is clear from (\ref{m75}) that at $g_A=g_B$ we have $A=B$, i.e. $G_s=0$. So in massless (2+1)-D 
{\it pure} Thirring model (without  $G_s$ coupling) the ground state is indeed a mixture of CSB and $\cP$-breaking
phases, i.e. this state can be imagined as a space filled with one of the above phases, in which bubbles of another 
phase can exist.
\begin{figure}
\hspace{-1cm}\includegraphics[width=0.4\textwidth]{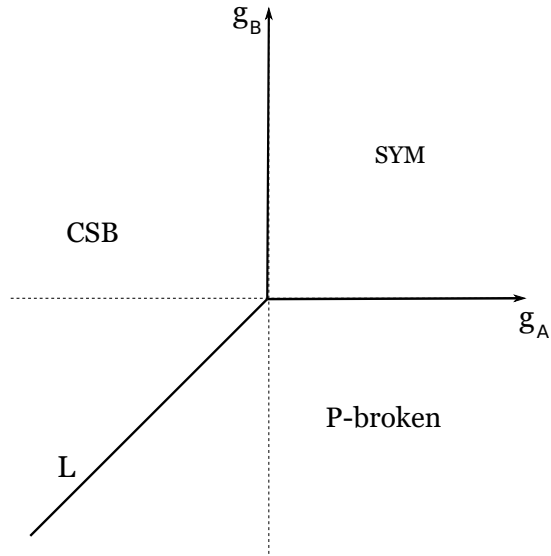}
 \caption{ $(g_A,g_B)$-phase portrait of the generalized Thirring model. Here, we use the notations CSB and 
 ''P-broken`` for chiral symmetry breaking and parity $\cP$-breaking phases, respectively. The notation SYM means the symmetrical
 phase without mass generation. On the stright line $L$ $g_A=g_B$.}
\end{figure}

\section{Phase portrait in terms of dimensionless bare couplings}

Finally, let us look at the properties of the model (\ref{t1}) from renormalization group point of view, i.e. try to 
find a position of its UV-stable fixed point as well as depict its phase portrait, in contrast to 
phase diagram of Fig. 1, in terms of some dimensionless parameters. To this aim, we should attract some dimensionless bare 
quantities and then find the zeros of the corresponding Callan-Simanzik $\beta$ functions. In our case, it is most 
convenient to deal with the bare quantities $A$ and $B$ (\ref{m72}). We have shown that in the Hartree-Fock 
approximation the model is renormalizable if these couplings behave vs $\Lambda$ as it is shown in Eqs. (\ref{m73}). 
Taking into account such a dependence of $A$ and $B$ on $\Lambda$, we can now determine the following dimensionless 
bare quantities $\lambda\equiv\lambda(\Lambda)=\Lambda A(\Lambda)$ and $\mu\equiv\mu(\Lambda)=\Lambda B(\Lambda)$, 
and the corresponding Callan-Simanzik $\beta$ functions (for definition, see, e.g., in Sec. 2.7 of Ref. 
\cite{Rosenstein}) 
\begin{eqnarray}
\beta_A (\lambda)\equiv\Lambda\partial\lambda/\partial\Lambda=2\lambda (\frac 12-\lambda),~~~\beta_B (\mu)\equiv
\Lambda\partial\mu/\partial\Lambda=2\mu (\frac 12-\mu).
 \label{t78}
\end{eqnarray}
Due to the structure (\ref{t78}) of these Callan-Simanzik $\beta$ functions, it is clear that both $\lambda(\Lambda)$ 
and $\mu(\Lambda)$ tend to $1/2$ when $\Lambda\to\infty$. It means that in the $(\mu,\lambda)$-plane there exists an 
UV-stable fixed point with coordinates $(1/2,1/2)$. \footnote{Note that this conclusion also follows directly from
Eq. (\ref{m740}) when $\Lambda\to\infty$.} Then, taking into account the relations (\ref{m740}), it is also possible 
to establish that at sufficiently high values of $\Lambda$
\begin{eqnarray}
\lambda -1/2 = -\frac{\pi g_A}{8\Lambda}+\cdots,~~~\mu -1/2 = -\frac{\pi g_B}{8\Lambda}+\cdots
 \label{t79}
\end{eqnarray} 
It follows from Eqs. (\ref{t79}) that at $\lambda<1/2$ and $\mu<1/2$ we have both $g_A>0$ and $g_B>0$. According to 
a phase portrait of Fig. 1, it corresponds to symmetrical phase of the model. It means that 
in the region $\{(\mu,\lambda):~\lambda<1/2,\mu<1/2\}$ of the $(\mu,\lambda)$-plane the symmetric phase 
is arranged. In a similar way, using the relations (\ref{t79}) between dimensional $g_A,g_B$ and dimensionless
$\lambda,\mu$ couplings and taking into account phase diagram of Fig. 1, one can draw the phase portrait of the 
model in terms of $\lambda$ and $\mu$ in some vicinity of the UV-stable fixed point with coordinates $(1/2,1/2)$ (see in Fig. 2). 

Alternatively, it is also possible to remake the phase portrait of Fig. 2 of the model in terms of other, more 
natural and physically acceptable dimensionless coupling constants, $g_s\equiv\Lambda G_s$ and $g_v\equiv\Lambda G_v$. Due to 
Eqs. (\ref{m72}), they are connected with $\lambda$ and $\mu$ by the relations
\begin{eqnarray}
4N\pi^2\lambda = -g_s+3g_v,~~4N\pi^2\mu = (4N-1)g_s+3g_v.
 \label{t80}
\end{eqnarray} 
It is clear from Eq.  (\ref{t80}) that the lines $\mu=\lambda$, $\lambda=1/2$ and $\mu=1/2$ of Fig. 2 transforms, 
respectively, to the lines $g_s=0$, $\it l_1$ and $\it l_2$ of the $(g_s,g_v)$-plane, where  
\begin{eqnarray}
{\it l_1}:~ g_v=\frac 13 g_s+\frac{2N\pi^2}{3},~~{\it l_2}:~ g_v=-g_s\frac{4N-1}{3}+\frac{2N\pi^2}{3}.
 \label{t81}
\end{eqnarray} 
These lines intersect in the UV-fixed point with coordinates $(g_s=0,g_v=g_v^*)$, where $g_v^*=\frac{2N\pi^2}{3}$. 
So in Fig. 3 the $(g_s,g_v)$-phase portrait of the model is presented in some neighborhood of this UV-fixed point.

It follows from the phase diagram of Fig. 3 that in the framework of the HF approximation the initial symmetry of the 
generalized Thirring model (\ref{t1}) can be broken dynamically at arbitrary fixed value of $N$. Namely, suppose that 
$g_v\gtrsim g_v^*$. Then at sufficiently small and positive values of $g_s$ the $\cP$-breaking phase is realized in 
the model
and the Haldane fermion mass is dynamically generated. However, when $g_s$ is small and negative, then fermions 
aquire dynamically the Dirac mass, and in this case both chiral and $U(2N)$ symmetries of the model are broken 
spontaneously. In the particular case when $g_s=0$, but $g_v>g_v^*$ there is a coexistence of these phases 
(this situation is realized in the original (2+1)-D Thirring model with only one nonzero coupling $G_v$). 

It is clear from Eq. (\ref{t81}) that stright lines $l_1$ and $l_2$ intersect the $g_s$ axis of Fig. 3 in the 
points $g_s^*$ and $g_s^{**}$, respectively, where $g_s^*=-2N\pi^2$ and $g_s^{**}=\frac{2N\pi^2}{4N-1}$. 
Hence if $N\to\infty$, then we have $g_v^*\to\infty$, $g_s^*\to-\infty$ and $g_s^{**}\to\pi^2/2$. As a result, we have 
in this case the expansion of the symmetrical phase over the whole region of the $(g_s,g_v)$ plane, such that 
$g_s<\pi^2/2$ (of course, it contains the $g_v$ axis). This fact corresponds to the absence of symmetry 
breaking effects in the generalized massless (2+1)-D Thirring model if it is studied by the HF method at $N\to\infty$
(and for sufficiently small values of $g_s<\pi^2/2$). The similar property of the pure (2+1)-D massless Thirring model 
is observed when it is 
investigated in the framework of the leading order of the large-$N$ technique (see, e.g., in Ref. \cite{Hong}).
\begin{figure}
 \includegraphics[width=0.4\textwidth]{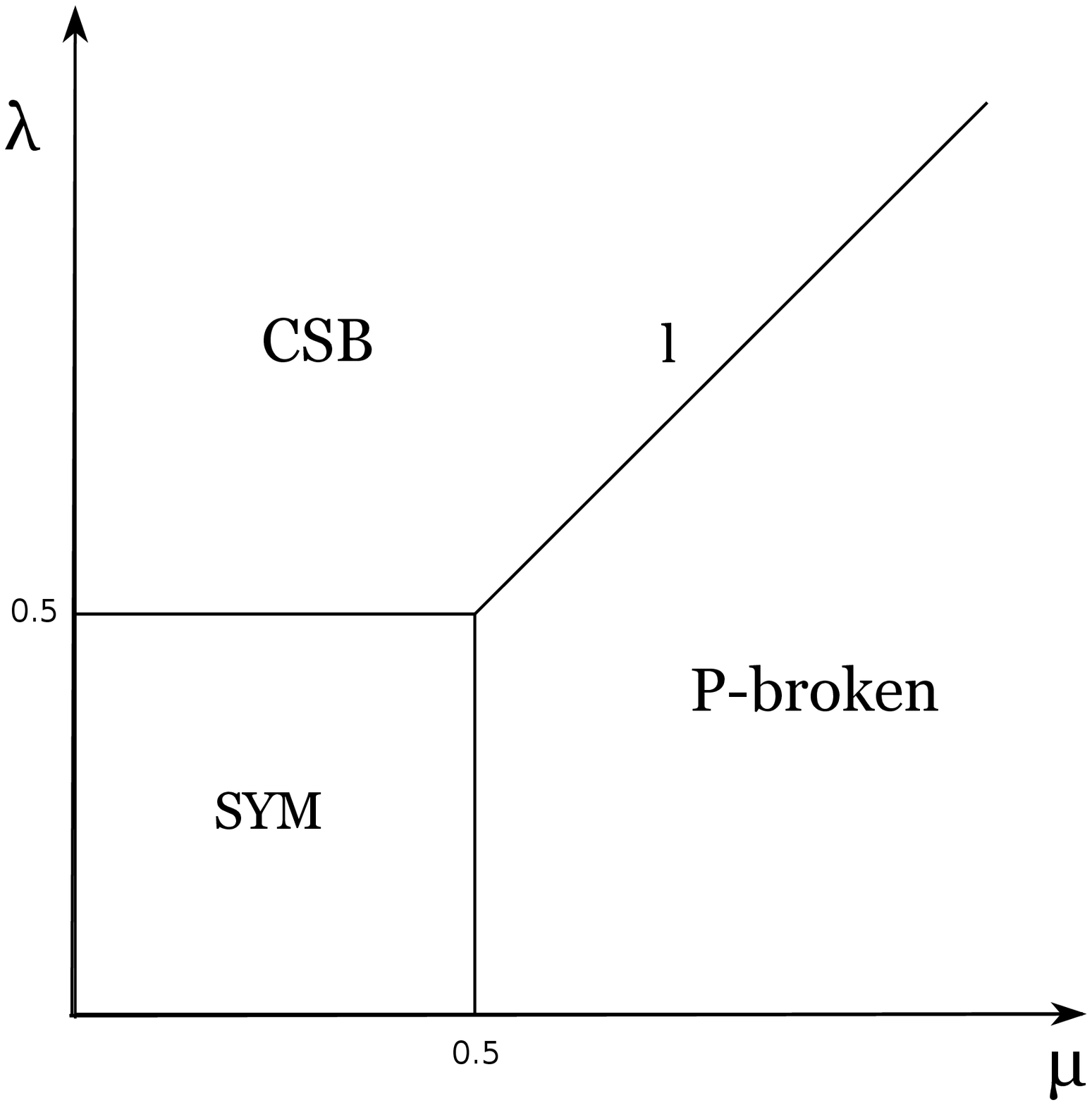}
 \hfill
 \includegraphics[width=0.5\textwidth]{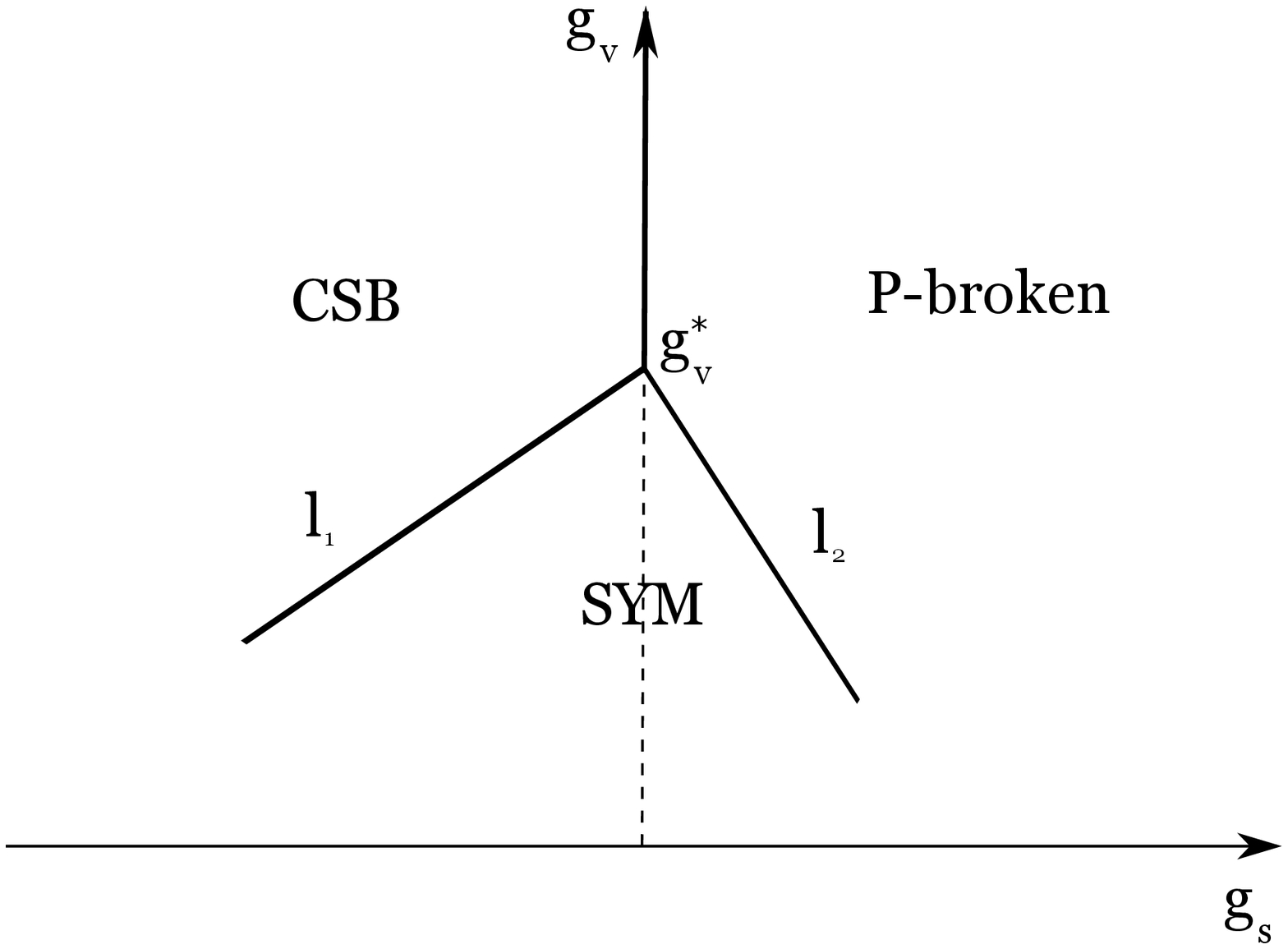}\\
\parbox[t]{0.45\textwidth}{ \caption{Phase portrait of the model in terms of dimensionless bare couplings 
$\mu$ and $\lambda$ defined in the text before Eq. (\ref{t78}). The point $(0.5,0.5)$ of the $(\mu,\lambda)$ plane is the UV-fixed point. the line
$l$ is defined by equation $\lambda=\mu$. Other notations are introduced in Fig. 1.}
 }\hfill
 \parbox[t]{0.45\textwidth}{
\caption{Phase portrait of the model in terms of dimensionless bare couplings 
$g_s$ and $g_v$ defined in the text before Eq. (\ref{t80}). The point $(0,g_v^*)$ (where $g_v^*=2N\pi^2/3$) of the $(g_s,g_v)$ plane is the 
UV-fixed point. The lines $l_1$ and $l_2$ are defined in Eq. (\ref{t81}). Other notations are introduced in Fig. 1.} }
\end{figure}

\section{Summary and conclusions}

In this work, the phase structure of the massless (2+1)-D generalized Thirring model (\ref{t1}), in which fermions are 
four-component, is studied by the Hartree-Fock method. The method is based on the 
Cornwall-Jackiw-Tomboulis effective action for composite operators (see in Section 2) calculated up to the first 
order in the coupling constants \cite{CJT}. 
In our opinion, one of the advantages of this CJT approach is the possibility to study the phase structure of any 
four-fermionic quantum field theory model without introducing auxiliary scalar (as it is often done in the case 
of the Gross-Neveu models) or vector fields -- in the case of Thirring model, etc. 

Prior to this, the HF approach was not used when considering 
the properties of the pure Thirring model, i.e. when $G_s=0$ in Eq. (\ref{t1}). At the same time, other approaches 
($1/N$-expansion, variational method, etc.) gave contradictory information regarding the structure of the ground state 
($\equiv$ of the vacuum) of the model. 
For example, some papers predict the dynamic generation of the Dirac mass 
$m_D\overline \Psi_k\Psi_k$ and appearance of a phase with broken chiral symmetry \cite{Hong,Hyun,Itoh}. 
In others, the ground state is characterized by $\cP$-parity violation and the appearance of Haldane mass 
$m_H\overline \Psi_k \tau\Psi_k$ \cite{Rossini,Ahn} for fermions (about other inconsistences see, e.g., in Refs. 
\cite{Janssen,Wipf,Wipf2,Hands2}). 

Using the HF approach, we were able to show that there is no contradiction between the above mentioned results, 
since in fact the vacuum of the (2+1)-D pure Thirring model
is really a mixed state in which these two phases coexist. In other words, in a part of the two-dimensional space, 
the Dirac mass is dynamically generated for fermions, and chiral symmetry is spontaneously broken in this region.
At the same time, the region of this phase can border on areas of another phase, in which the Haldane mass is 
generated, and $\cP$ parity is spontaneously broken. During the transition from one phase to another, a first-order 
phase transition occurs in the system (see the last paragraph of the Section III). Moreover, it is clear that in the 
framework of the HF approach to the pure (2+1)-D Thirring model (\ref{t1}) the dynamical mass generation comes at any 
finite $N$ only from the Fock term of Eq. (\ref{n043}). At $G_s=0$ the Hartree term makes no contribution to the 
regularized HF equation (\ref{m7}). Since the Fock term is proportional to $1/N$, we see the absence of dynamical 
mass generation in the limit $N\to\infty$. The same result was obtained in the leading order of the $1/N$-expansion 
approach to this model \cite{Hong}. 

Returning to the results obtained when considering the properties of the generalized Thirring model (\ref{t1})
by the HF method, we note that renormalized (i.e. finite) expressions both for the effective potential (\ref{m77})
and for the HF equation (\ref{n043}) itself can only be obtained for a well-defined behavior (\ref{m75}) of the bare 
coupling constants $G_s(\Lambda)$ and $G_v(\Lambda)$ vs cutoff parameter $\Lambda$, in which two finite 
(and renormalization group invariant) constants $g_A$ and $g_B$ appear. In this case, for arbitrary fixed values of 
$g_A$ and $g_B$, only one of the phases is realized in the model, symmetric, $\cP$-breaking, or a phase with chiral 
symmetry breaking (the last two phases are characterized by dynamic appearance of the Haldane  or Dirac mass, 
respectively), and the $(g_A,g_B)$-phase portrait of the model is shown in Fig. 1. 

Then, using the dimensionless couplings $g_s\equiv\Lambda G_s(\Lambda)$ and $g_v\equiv\Lambda G_v(\Lambda)$, 
we have shown that generalized Thirring model (\ref{t1}) is characterized by non-trivial UV stability. 
It means that in the $(g_s,g_v)$-plane there exists a so-called UV-stable fixed point $(g_s=0,g_v=g_v^*)$, 
where $g_v^*=\frac{2N\pi^2}{3}$, such that in the limit $\Lambda\to\infty$ we have  $(g_s,g_v)\to (0,g_v^*)$. 
Phase portrait of the model in some neighborhood of this UV-stable fixed point is given in Fig. 3.

It follows from Fig. 3 that at each fixed $N$, when UV-fixed point is finite, dynamical generation of fermion mass,
Dirac or Haldane, is possible in the generalized Thirring model (see discussion at the end of section IV). However, if
$N\to\infty$, then UV-fixed point tends to $\infty$ along the $g_v$ axis, and for arbitrary fixed 
values of dimensionless couplings $g_s$ and $g_v$ (when $g_s$ is a rather small) the point $(g_s,g_v)$ lies in the 
region corresponding to symmetrical phase, i.e. the dynamical generation of any fermion mass is absent 
(similar to the results of Refs. \cite{Hong,Hyun} obtained in the pure (2+1)-D Thirring model). 

Finally, two remarks are in order. First, in the recent study of the (2+1)-D Gross-Neveu model by HF method 
\cite{Tamaz1} just the Hartree term gives the main contribution to the dynamic generation of the fermion mass, 
i.e. to the effect that is also observed in the leading order of the large-$N$ approximation \cite{Rosenstein} 
(the contribution of the Fock term in this case is not so significant). In contrast, the present investigation of the 
generalized (2+1)-D Thirring model (\ref{t1}) by the HF method shows that at $G_s\ne 0$ the Fock terms of the 
stationary equation (\ref{n043}) play a more important role in the dynamical generation of a fermion mass
(and at $G_s=0$, i.e., in the original Thirring model, the appearance of this effect is entirely due to the 
Fock terms). Since the Fock terms are proportional to $1/N$, it might seem that HF approach to fermion self-energy 
is equivalent to its study in the framework of first two orders of large-$N$ expansion. Indeed, as it is discussed 
in Ref. \cite{Klevansky}, taking into account the Hartree terms in equations of the type (\ref{n043}) is equivalent 
to considering the properties of the fermion propagator in the leading order of the $1/N$ expansion. However, there 
are a lot of diagrams that are of $1/N$ order, but which lie outside the scope of the HF consideration 
and are not described by Fock terms \cite{Klevansky}. Thus, it is because of the presence of the Fock terms 
that the difference between the HF and large-$N$ methods appears.

Second, HF method is a kind of the well-known mean-field approach widely used in both field theory and many-particle 
physics. And, of course, its scope is limited. For example, the HF approach to (2+1)-D Thirring model predicts 
dynamical symmetry breaking at any fixed finite value of $N$. However, if the model is investigated by other 
and more sophisticated nonperturbative methods (such as the fuctional renormalization group, the Dyson-Schwinger 
method, and, especially, the lattice approach, which is based on the first principles of quantum field theory), the 
acceptable values of $N<N_c$ are rather small (the value of $N_c$ is discussed, e.g., in the recent review 
\cite{Wipf2}). In such cases, in our opinion, it is possible to improve the results of the HF method by going 
beyond the mean-field approach using the next orders over couplings $G_{s,v}$ in the CJT effective action 
$\Gamma(S)$ (\ref{0360}).

\section{ACKNOWLEDGMENTS}

R.N.Z. is grateful for support of the Foundation for the Advancement of Theoretical Physics and 
Mathematics BASIS.

\appendix
\section{Algebra of the $\gamma$ matrices in the case of $SO(2,1)$ group}
\label{ApC}

The two-dimensional irreducible representation of the (2+1)-dimensional
Lorentz group $SO(2,1)$ is realized by the following $2\times 2$
$\tilde\gamma$-matrices:
\begin{eqnarray}
\tilde\gamma^0=\sigma_3=
\left (\begin{array}{cc}
1 & 0\\
0 &-1
\end{array}\right ),\,\,
\tilde\gamma^1=i\sigma_1=
\left (\begin{array}{cc}
0 & i\\
i &0
\end{array}\right ),\,\,
\tilde\gamma^2=i\sigma_2=
\left (\begin{array}{cc}
0 & 1\\
-1 &0
\end{array}\right ),
\label{C1}
\end{eqnarray}
acting on two-component Dirac spinors. They have the properties:
\begin{eqnarray}
Tr(\tilde\gamma^{\mu}\tilde\gamma^{\nu})=2g^{\mu\nu};~~
[\tilde\gamma^{\mu},\tilde\gamma^{\nu}]=-2i\varepsilon^{\mu\nu\alpha}
\tilde\gamma_{\alpha};~
~\tilde\gamma^{\mu}\tilde\gamma^{\nu}=-i\varepsilon^{\mu\nu\alpha}
\tilde\gamma_{\alpha}+g^{\mu\nu},
\label{C2}
\end{eqnarray}
where $g^{\mu\nu}=g_{\mu\nu}=diag(1,-1,-1),
~\tilde\gamma_{\alpha}=g_{\alpha\beta}\tilde\gamma^{\beta},~
\varepsilon^{012}=1$.
There is also the relation:
\begin{eqnarray}
Tr(\tilde\gamma^{\mu}\tilde\gamma^{\nu}\tilde\gamma^{\alpha})=
-2i\varepsilon^{\mu\nu\alpha}.
\label{C3}
\end{eqnarray}
Note that the definition of chiral symmetry is slightly unusual in
(2+1)-dimensions (spin is here a pseudoscalar rather than a (axial)
vector). The formal reason is simply that there exists no other 
$2\times 2$ matrix anticommuting with the Dirac matrices $\tilde\gamma^{\nu}$
which would allow the introduction of a $\gamma^5$-matrix in the
irreducible representation. The important concept of 'chiral'
symmetries  and their breakdown by mass terms can nevertheless be
realized also in the framework of (2+1)-dimensional quantum field
theories by considering a four-component reducible representation
for Dirac fields. In this case the Dirac spinors $\psi$ have the
following form:
\begin{eqnarray}
\psi(x)=
\left (\begin{array}{cc}
\tilde\psi_{1}(x)\\
\tilde\psi_{2}(x)
\end{array}\right ),
\label{C4}
\end{eqnarray}
with $\tilde\psi_1,\tilde\psi_2$ being two-component spinors.
In the reducible four-dimensional spinor representation one deals
with 4$\times$4 $\gamma$-matrices:
$\gamma^\mu=diag(\tilde\gamma^\mu,-\tilde\gamma^\mu)$, where
$\tilde\gamma^\mu$ are given in (\ref{C1}) (This particular reducible representation for 
$\gamma$-matrices is used, e.g., in Ref. \cite{Appelquist}). One can easily show, that
($\mu,\nu=0,1,2$):
\begin{eqnarray}
&&Tr(\gamma^\mu\gamma^\nu)=4g^{\mu\nu};~~
\gamma^\mu\gamma^\nu=\sigma^{\mu\nu}+g^{\mu\nu};~~\nonumber\\
&&\sigma^{\mu\nu}=\frac{1}{2}[\gamma^\mu,\gamma^\nu]
=diag(-i\varepsilon^{\mu\nu\alpha}\tilde\gamma_\alpha,
-i\varepsilon^{\mu\nu\alpha}\tilde\gamma_\alpha).
\label{C5}
\end{eqnarray}
In addition to the  Dirac matrices $\gamma^\mu~~(\mu=0,1,2)$ there
exist two other matrices, $\gamma^3$ and $\gamma^5$, which anticommute
with all $\gamma^\mu~~(\mu=0,1,2)$ and with themselves
\begin{eqnarray}
\gamma^3=
\left (\begin{array}{cc}
0~,& I\\
I~,& 0
\end{array}\right ),\,
\gamma^5=\gamma^0\gamma^1\gamma^2\gamma^3=
i\left (\begin{array}{cc}
0~,& -I\\
I~,& 0
\end{array}\right ),\,\,\tau=-i\gamma^3\gamma^5=
\left (\begin{array}{cc}
I~,& 0\\
0~,& -I
\end{array}\right )
\label{C6}
\end{eqnarray}
with  $I$ being the unit $2\times 2$ matrix.

\section{Calculation of the $\Gamma (S)$ up to a first order in $G_{s,v}$: Hartree-Fock approximation}
\label{ApB}
\subsection{The case $G_{s,v}=0$}

In this case $\exp\left (iI_{int}\left (-i\frac{\delta}{\delta K}\right )\right )=1$, so we 
have from Eqs. (\ref{36})-(\ref{036})
\begin{eqnarray}
\exp(iNW(K))&=&\exp \Big [N{\rm Tr}\ln \big (D(x,y)+K(x,y)\big )\Big ] 
\nonumber\\
\Longrightarrow W(K)&=&-i{\rm Tr}\ln \big (D(x,y)+K(x,y)\big ).
 \label{38}
\end{eqnarray}
Now, using a well-known relation (see, e.g., Eq. (11.101) of Ref. \cite{peskin2}), 
\begin{eqnarray}
\frac{\partial}{\partial\alpha}{\rm Tr}\ln M(\alpha)=
{\rm Tr}\left\{ M^{-1}\frac{\partial M}{\partial\alpha}\right\},\label{A50}
\end{eqnarray}
where $M\equiv M(\alpha)$ is a matrix, 
we have from Eqs. (\ref{37}) and  (\ref{38})
$$
S^\alpha_\beta(x,y)=\frac{\delta W(K)}{\delta K_\alpha^\beta(y,x)}=
-i\int d^3sd^3t\sum_{\mu\nu}\Big [ \big (D+K\big )^{-1}\Big ]^\mu_\nu(s,t)
\frac{\delta K^\nu_\mu(t,s)}{\delta K_\alpha^\beta(y,x)}
$$\vspace{-0.5cm}
\begin{eqnarray}
=-i\int d^3sd^3t\sum_{\mu\nu}\Big [ \big (D+K\big )^{-1}\Big ]^\mu_\nu(s,t)\delta^3 (t-y)
\delta^3 (s-x) \delta_{\nu\beta}\delta_{\mu\alpha}
=-i\Big [ \big (D+K\big )^{-1}\Big ]^\alpha_\beta(x,y).
 \label{380}
\end{eqnarray}
Solving this equation with respect to $K$, we obtain
\begin{eqnarray}
K=-iS^{-1}-D. \label{038}
\end{eqnarray}
Finally, after substituting the relation (\ref{038}) into Eq. (\ref{38}) and taking into account 
the definition (\ref{0360}) of the CJT effective action $\Gamma(S)$, we have (omitting independent 
of $S$ terms) for it the following expression at $G=0$,
\begin{eqnarray}
\Gamma (S)=-i{\rm Tr}\ln \big (-iS^{-1}\big )+\int d^3xd^3y S^\alpha_\beta(x,y)D_\alpha^\beta(y,x).
 \label{0380}
\end{eqnarray}
Starting from the CJT effective action (\ref{0380}), it is possible to obtain the stationary 
equation (see Eq. (\ref{0370})) for the genuine spinor propagator $S$ of the generalized (2+1)-D Thirring model at 
$G_{s,v}=0$. Taking into account the relation (\ref{A50}), it can be presented in the following form
\begin{eqnarray}
0&=&i\int d^3sd^3t\sum_{\mu\nu}\Big [S^{-1}\Big ]^\mu_\nu(s,t)\frac{\delta S^\nu_\mu(t,s)}{\delta S^\alpha_\beta(x,y)}+D_\alpha^\beta(y,x)
=i\Big [S^{-1}\Big ]^\beta_\alpha(y,x)+D_\alpha^\beta(y,x),
 \label{39}
\end{eqnarray}
where a trivial relation $\frac{\delta S^\nu_\mu(t,s)}{\delta S^\alpha_\beta(x,y)}=
\delta^3 (t-x)\delta^3 (s-y) \delta_{\nu\alpha}\delta_{\mu\beta}$ 
is taken into consideration. Hence, in the absence of interaction in the generalized Thirring model (\ref{t1}), 
i.e. at $G_{s,v}=0$, the stable and stationary form of the propagator is the following, $S=-iD^{-1}$, where $D$ is 
presented in Eq. (\ref{360}).

\subsection{CJT effective action in the first order in coupling constants}

In this case the functional $W(K)$ (\ref{036}) looks like (here and below we use the 
definition $\Delta\equiv D+K$)
\begin{eqnarray}
&&\exp(iNW(K))=\Big (1+iI_{v}\Big (-i\frac{\delta}{\delta K}\Big )+iI_{s}\Big (-i\frac{\delta}{\delta K}\Big )\Big )
\exp \Big (N{\rm Tr}\ln \Delta\Big )\nonumber\\
&=&\Big\{1+i\frac {G_v}{2N}\int d^3sd^3td^3ud^3v\delta^3 (s-t)\delta^3 (t-u)\delta^3 (u-v) 
(\gamma^\rho)_\alpha^\beta\frac{\delta}{\delta K^\beta_\alpha (s,t)} 
(\gamma_\rho)_\mu^\nu\frac{\delta}{\delta K^\nu_\mu (u,v)}\nonumber\\
&&-i\frac {G_s}{2N}\int d^3sd^3td^3ud^3v\delta^3 (s-t)\delta^3 (t-u)\delta^3 (u-v) 
\tau_\alpha^\beta\frac{\delta}{\delta K^\beta_\alpha (s,t)} 
\tau_\mu^\nu\frac{\delta}{\delta K^\nu_\mu (u,v)}
\Big\}\exp \Big (N{\rm Tr}\ln \Delta\Big ).
\label{039}
\end{eqnarray}
In the following, two relations are needed,
\begin{eqnarray}
\frac{\delta {\rm Tr} \ln\Delta}{\delta K_\mu^\nu(u,v)}=\Big ( \Delta^{-1}\Big )^\mu_\nu(v,u),
 \label{390}
\end{eqnarray}
which is a consequence of Eq. (\ref{380}) or Eq. (\ref{A50}), and
\begin{eqnarray}
\frac{\delta}{\delta K_\alpha^\beta(s,t)}\Big ( \Delta^{-1}\Big )^\mu_\nu(v,u)=
-\int d^3v'd^3u'\sum_{\mu',\nu'}\Big ( \Delta^{-1}\Big )^\mu_{\mu'}(v,v')\frac{\delta 
\Delta^{\mu'}_{\nu'}(v',u')}{\delta K_\alpha^\beta(s,t)}\Big (
\Delta^{-1}\Big )_\nu^{\nu'}(u',u). \label{0390}
\end{eqnarray}
Note that Eq. (\ref{0390}) follows from a rather general formula (11.94) of Ref. \cite{peskin2}. 
Taking into account in Eq. (\ref{0390}) that 
$\frac{\delta \Delta^{\mu'}_{\nu'}(v',u')}{\delta K_\alpha^\beta(s,t)}=\delta^3 
(v'-s)\delta^3 (u'-t)\delta^{\mu'\beta}\delta_{\nu'\alpha}$, we have 
\begin{eqnarray}
\frac{\delta}{\delta K_\alpha^\beta(s,t)}\Big ( \Delta^{-1}\Big )^\mu_\nu(v,u)=
-\Big ( \Delta^{-1}\Big )^\mu_{\beta}(v,s)\Big ( \Delta^{-1}\Big )_\nu^{\alpha}(t,u).\label{040}
\end{eqnarray}
Applying the relations (\ref{390}) and (\ref{040}) in Eq. (\ref{039}), we obtain
\begin{eqnarray}
&&\exp(iNW(K))=\Big\{1
+i\frac{G_vN}2\int d^3s ~\left ({\rm tr}\big [\gamma^\rho\Delta^{-1}(s,s)
\big ]{\rm tr}\big [\gamma_\rho\Delta^{-1}(s,s)\big ]\right )-i\frac{G_v}2
\int d^3s~ {\rm tr}\Big [\gamma^\rho\Delta^{-1}(s,s)\gamma_\rho\Delta^{-1}(s,s)\Big ]
\nonumber\\&&
-i\frac{G_sN}2\int d^3s ~\left ({\rm tr}\big [\tau\Delta^{-1}(s,s)\big ]\right )^2
+i\frac{G_s}2
\int d^3s~ {\rm tr}\Big [\tau\Delta^{-1}(s,s)\tau\Delta^{-1}(s,s)\Big ]
\Big\}\exp \Big (N{\rm Tr}\ln \Delta\Big ),
 \label{40}
\end{eqnarray}
where ${\rm tr}$ means the trace operation only in the spinor space. It follows from Eq. (\ref{40})
that up to a first order in $G_{v,s}$
\begin{eqnarray}
W(K)&=&-i{\rm Tr}\ln \Delta-\frac{G_s}2\int d^3s \left ({\rm tr}\big [\tau\Delta^{-1}(s,s)
\big ]\right )^2 
+\frac{G_s}{2N}\int d^3s~ {\rm tr}\Big [\tau\Delta^{-1}(s,s)\tau\Delta^{-1}(s,s)\Big ]
\nonumber\\&&
+\frac{G_v}2\int d^3s \left ({\rm tr}\big [\gamma^\rho\Delta^{-1}(s,s)
\big ] {\rm tr}\big [\gamma_\rho\Delta^{-1}(s,s)\big ]\right ) 
-\frac{G_v}{2N}\int d^3s~ {\rm tr}\Big [\gamma^\rho\Delta^{-1}(s,s)\gamma_\rho\Delta^{-1}(s,s)\Big ].
 \label{400}
\end{eqnarray}
To find the effective action $\Gamma(S)$ in the first order of $G_v$ and $G_s$ (or in the HF approximation),
we must use in Eq. (\ref{0360}), as well as in Eq. (\ref{37}), the expression (\ref{400}) for $W(K)$. In particular,
it follows from Eqs. (\ref{37}) and (\ref{400}) that
\begin{eqnarray}
S^\alpha_\beta(x,y)\equiv\frac{\delta W(K)}{\delta K_\alpha^\beta(y,x)}&=&
-i\Big ( \Delta^{-1}\Big )_\beta^\alpha(x,y)+G_s\int d^3s~
\Big [\Delta^{-1}(x,s)\tau \Delta^{-1}(s,y)\Big ]^\alpha_\beta{\rm tr}\big [\tau\Delta^{-1}(s,s)
\big ]\nonumber\\&&
-\frac{G_s}{N}\int d^3s~ \Big [\Delta^{-1}(x,s)\tau\Delta^{-1}(s,s)
\tau\Delta^{-1}(s,y)\Big ]^\alpha_\beta\nonumber\\&&
-G_v\int d^3s~
\Big [\Delta^{-1}(x,s)\gamma^\rho \Delta^{-1}(s,y)\Big ]^\alpha_\beta{\rm tr}\big [\gamma_\rho\Delta^{-1}(s,s)
\big ]\nonumber\\&&
+\frac{G_v}{N}\int d^3s~ \Big [\Delta^{-1}(x,s)\gamma^\rho\Delta^{-1}(s,s)
\gamma_\rho\Delta^{-1}(s,y)\Big ]^\alpha_\beta,
 \label{0400}
\end{eqnarray}
where the relation (\ref{040}) was applied. Now, the next problem is to express the bilocal 
sourse $K$ as a function(al) of $S$ with a help of Eq. (\ref{0400}). We will use the perturbation 
approarch over the coupling constants $G_{s,v}$, i.e., will suppose that the solution of Eq. (\ref{0400}) has the form
\begin{eqnarray}
K(S)=K_0+\delta K,
 \label{41}
\end{eqnarray}
where $\delta K\sim G_{s,v}$ and $K_0$ is the solution of Eq. (\ref{0400}) at $G_{s,v}=0$, and it is given 
in Eq. (\ref{038}), i.e., $K_0= -iS^{-1}-D$. Recall that $\Delta^{-1}$ in Eq. (\ref{0400}) is 
indeed a functional of $K$, i.e., $\Delta^{-1}\equiv \Delta^{-1}(K)$. So, let us expand this 
quantity in a Taylor series around $K_0$ up to a first order in a small perturbation 
$\delta K$ of Eq. (\ref{41}),
\begin{eqnarray}
\Big ( \Delta^{-1}(K)\Big )_\beta^\alpha(x,y)=\Big ( \Delta^{-1}(K_0)\Big )_\beta^\alpha(x,y)+\int d^3ud^3v~\delta K^\nu_\mu(u,v)\frac{\delta \Big ( \Delta^{-1}(K)\Big )_\beta^\alpha(x,y)}{\delta K^\nu_\mu(u,v)}\Big |_{K=K_0}+\cdots.
 \label{041}
\end{eqnarray}
Taking into account in Eq. (\ref{041}) the derivative rule (\ref{040}) as well as the trivial 
relation $\Big (\Delta^{-1}(K_0)\Big )_\beta^\alpha(x,y)=iS_\beta ^\alpha(x,y)$, we obtain 
\begin{eqnarray}
\Big ( \Delta^{-1}(K)\Big )_\beta^\alpha(x,y)=iS_\beta ^\alpha(x,y)+\int d^3ud^3v~S^\alpha_\nu (x,u)\delta K^\nu_\mu(u,v)S^\mu_\beta (v,y)+\cdots.
\label{410}
\end{eqnarray}
After a substitution of the relation (\ref{410}) instead of a first term in the right hand side of Eq. (\ref{0400}) and replacing all $\Delta^{-1}$ in other terms of Eq. (\ref{0400}) by $iS$, we find the following equation on the quantity
$\delta K$
\begin{eqnarray}
&&\int d^3ud^3v~S^\alpha_\nu (x,u)\delta K^\nu_\mu(u,v)S^\mu_\beta (v,y)=
-G_s\int d^3s~\big [S(x,s)\tau S(s,y)\big ]_\beta^{\alpha}~{\rm tr}\big [\tau S(s,s)\big ] 
\nonumber\\
&+&\frac{G_s}{N}\int d^3s~\big [S(x,s)\tau S(s,s)\tau S(s,y)\big ]_\beta^{\alpha}
+G_v\int d^3s~\big [S(x,s)\gamma^\rho S(s,y)\big ]_\beta^{\alpha}~{\rm tr}
\big [\gamma_\rho S(s,s)\big ]\nonumber\\
&-&\frac{G_v}{N}\int d^3s~\big [S(x,s)\gamma^\rho S(s,s)\gamma_\rho S(s,y)\big ]_\beta^{\alpha}.
 \label{0410}
\end{eqnarray}
Its solution with respect to $\delta K$ has the following form
\begin{eqnarray}
\delta K^\alpha_\beta(x,y)&=&-G_s\tau^\alpha_\beta {\rm tr}\big [\tau S(x,x)\big ]\delta^3 (x-y)
+\frac{G_s}{N}\big [\tau S(x,x)\tau\big ]_\beta^{\alpha}\delta^3 (x-y)\nonumber\\
&+&G_v\big (\gamma^\rho\big )^\alpha_\beta {\rm tr}\big [\gamma_\rho S(x,x)\big ]\delta^3 (x-y)
-\frac{G_v}{N}\big [\gamma^\rho S(x,x)\gamma_\rho\big ]_\beta^{\alpha}\delta^3 (x-y).
 \label{42}
\end{eqnarray}
Bearing in mind this expression for $\delta K$ as well as that $K_0= -iS^{-1}-D$, 
we obtain, up to a first order in $G_{s,v}$, the solution $K(S)$ (\ref{41}) of the 
equation (\ref{0400}),
\begin{eqnarray}
K^\alpha_\beta(x,y)&=&-i\Big (S^{-1}\Big )^\alpha_\beta(x,y)-D^\alpha_\beta(x,y)
-G_s\tau^\alpha_\beta{\rm tr}\big [\tau S(x,x)\big ]\delta^3 (x-y)
+\frac{G_s}{N}\big [\tau S(x,x)\tau\big ]_\beta^{\alpha}\delta^3 (x-y)\nonumber\\
&+&G_v\big (\gamma^\rho\big )^\alpha_\beta{\rm tr}\big [\gamma_\rho S(x,x)\big ]\delta^3 (x-y)
-\frac{G_v}{N}\big [\gamma^\rho S(x,x)\gamma_\rho\big ]_\beta^{\alpha}\delta^3 (x-y).
\label{042}
\end{eqnarray}
It follows from Eq. (\ref{042}) that
\begin{eqnarray}
&&\Delta (K)^\alpha_\beta(x,y)\equiv K^\alpha_\beta(x,y)+D^\alpha_\beta(x,y)=
-i\Big (S^{-1}\Big )^\alpha_\beta(x,y)
-G_s\tau^\alpha_\beta{\rm tr}\big [\tau S(x,x)\big ]\delta^3 (x-y)
\nonumber\\&&+\frac{G_s}{N}\big [\tau S(x,x)\tau\big ]_\beta^{\alpha}\delta^3 (x-y)
+G_v\big (\gamma^\rho\big )^\alpha_\beta{\rm tr}\big [\gamma_\rho S(x,x)\big ]\delta^3 (x-y)
-\frac{G_v}{N}\big [\gamma^\rho S(x,x)\gamma_\rho\big ]_\beta^{\alpha}\delta^3 (x-y).
\label{043}
\end{eqnarray}
Now, it is clear from Eqs. (\ref{042}) and (\ref{043}) that  (also up to a first order in $G_{s,v}$)
\begin{eqnarray}
&&-i{\rm Tr}\ln \Delta=-i{\rm Tr}\ln\big (-iS^{-1}\big )-G_s\int d^3s \Big [{\rm tr}\big (
\tau S(s,s)\big )\Big ]^2 
+\frac{G_s}{N}\int d^3s~ {\rm tr}\Big [\tau S(s,s)\tau S(s,s)\Big ]\nonumber\\&&
+G_v\int d^3s~{\rm tr}\big [\gamma^\rho S(s,s)\big ]{\rm tr}\big [\gamma_\rho S(s,s)\big ]
-\frac{G_v}{N}\int d^3s~{\rm tr}\big [S(s,s)\gamma^\rho S(s,s)\gamma_\rho\big ],\nonumber\\
&&-\int d^3xd^3y S^\alpha_\beta(x,y)K_\alpha^\beta(y,x)=
\int d^3xd^3y S^\alpha_\beta(x,y)D_\alpha^\beta(y,x)+G_s\int d^3s \Big [{\rm tr}
\big (\tau S(s,s)\big )\Big ]^2\nonumber\\&& 
-\frac{G_s}{N}\int d^3s~ {\rm tr}\Big [\tau S(s,s)\tau S(s,s)\Big ]-
G_v\int d^3s~{\rm tr}\big [\gamma^\rho S(s,s)\big ]{\rm tr}\big [\gamma_\rho S(s,s)\big ]
\nonumber\\&&
+\frac{G_v}{N}\int d^3s~{\rm tr}\big [S(s,s)\gamma^\rho S(s,s)\gamma_\rho\big ]
 \label{045}
\end{eqnarray}
(the last equation is valid up to unessential and $S$-independent infinite constant). Finally, 
replacing all $\Delta^{-1}$ functions in the last four terms of Eq. (\ref{400}) for $W(K)$
by $iS$, and taking into account the relations (\ref{045}), we obtain in the first order in $G_{s,v}$ 
for the CJT effective action $\Gamma (S)$  (\ref{0360}) the expression (\ref{n420}) (where we 
also took into account the trivial relation, $\ln\big (-iS^{-1}\big )=-\ln\big (iS\big )$).

\end{document}